# Watermark Overwriting Attack on StegaStamp algorithm

*I. F. Serzhenko[1], L. A. Khaertdinova[2], M. A. Pautov[3,5], A. V. Antsiferova[3,4]*

[1]Moscow Institute of Physics and Technology
[2]Innopolis University
[3]Ivannikov Institute for System Programming of the Russian Academy of Sciences (ISP RAS)
[4]Moscow State University Artificial Intelligence Institute
[5]AIRI

With the growing prevalence of generative models capable of producing photorealistic images, the task of detecting AI-generated content and identifying its source has become increasingly critical. One approach to preserving the authenticity of generative content is digital watermarking, which involves the embedding of encoded signals within images to establish provenance or ownership. When implemented in an appropriate manner, the process of watermark extraction facilitates the unambiguous identification of content ownership. In order to maintain visual fidelity, it is particularly desirable to use imperceptible watermarks that evade human visual detection.

A crucial consideration in watermarking systems is their robustness against removal attacks. While many existing algorithms demonstrate resilience against conventional image manipulations, various watermark removal techniques continue to prove effective. The WAVES research group[1] has developed a comprehensive benchmark for evaluating watermark robustness, incorporating a diverse range of novel and realistic attacks, including traditional image distortions, image regeneration techniques, and adversarial attacks.

This study presents an attack methodology on the StegaStamp[2] watermarking algorithm, developed as part of participation in the international "NeurIPS2024 Competition: Erasing the Invisible." The competition framework was specifically designed to test the attack resilience of watermark embedding algorithms under rigorous evaluation conditions.

StegaStamp[2] is a steganographic algorithm that has been designed for reliable encoding and decoding of arbitrary bit strings as hyperlinks within digital images. The system has been developed to transform input images into embedded watermarks that remain virtually imperceptible to human observers. At its core, StegaStamp employs a deep neural network trained to perform encoding and decoding operations that maintain robustness against image perturbations resembling those encountered in real-world scenarios, including printing, scanning, and photographic capture. Furthermore, the algorithm demonstrates resilience against common image manipulations such as geometric distortions, cropping, compression artifacts, brightness/contrast adjustments, and similar transformations.

The StegaStamp workflow is initiated as follows: the algorithm accepts two parameters: an input image and a target hyperlink. Initially, the hyperlink is mapped to a unique bit string representation. The StegaStamp encoder then embeds this bit string into the host image, thereby producing an encoded output that maintains visual parity with the original. The watermarked image is then physically reproduced through the process of printing, followed by recapture via photography or scanning. The detection pipeline then processes the recaptured image through a trained neural network for preliminary enhancement before final decoding. The StegaStamp decoder ultimately extracts the embedded bit string, thereby enabling hyperlink reconstruction and access.

The proposed Watermark Overwriting Attack method has been demonstrated to be effective in the removal of embedded watermarks from images, while ensuring that visual quality is not significantly compromised. The attack algorithm operates through the following sequential steps, as illustrated in Figure 1:

1. Hidden Message Recovery: Extraction of the concealed binary message from watermarked images using the StegaStamp decoder.
2. Message Inversion: Bitwise inversion of the extracted binary message (converting all 0s to 1s and vice versa).
3. Message Re-embedding: Insertion of the inverted message into the original images using the StegaStamp encoder.

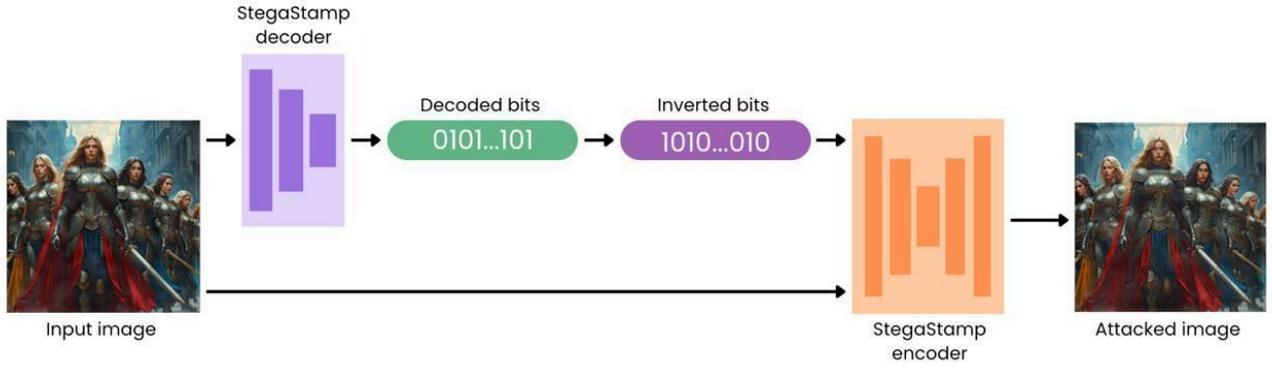

Figure 1. Watermark Overwriting Attack Method on StegaStamp algorithm

As demonstrated in Figure 2, a comparison is presented of an attacked image and a watermarked image, thereby illustrating the discrepancy between them.

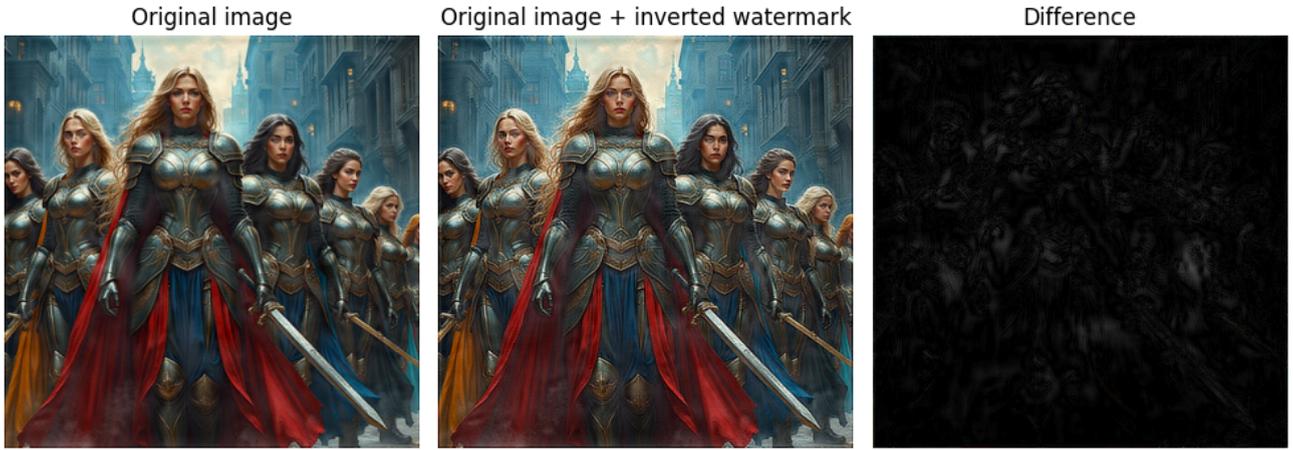

Figure 2. Attacked, watermarked images and difference between them

In the context of the NeurIPS competition framework, the performance of the watermark removal method employed was subjected to a rigorous evaluation. This assessment was conducted using the WAVES benchmark [1], a standard metric that quantifies the performance of watermark removal methods by assessing the trade-off between the efficacy of the method and the resulting image quality. The overall score was thus computed as $\sqrt{A^2 + Q^2}$, where Q represents the normalized image quality metric, and A denotes the watermark removal accuracy, measured as TPR@0.1%FPR.

The composite image quality metric Q incorporates eight distinct quality measures: Peak Signal-to-Noise Ratio (PSNR), Structural Similarity Index (SSIM), Normalized Mutual Information (NMI), Fréchet Inception Distance (FID), CLIPFID, Learned Perceptual Image Patch Similarity (LPIPS), Aesthetic score variation (ΔAesthetics), Artifact score variation (ΔArtifacts). The Q metric was calculated using the following weighted formula:

$$Q = 1.53 \times 10^{-3}\, \text{FID} + 5.07 \times 10^{-3}\, \text{CLIPFID} - 2.22 \times 10^{-3}\, \text{PSNR} - 1.13 \times 10^{-1}\, \text{SSIM} - 9.88 \times 10^{-2}\, \text{NMI} + 3.41 \times 10^{-1}\, \text{LPIPS} + 4.50 \times 10^{-2}\, \Delta\text{Aesthetics} - 1.44 \times 10^{-1}\, \Delta\text{Artifacts}$$

The team's solution has been awarded the 5th place out of the 65 participants in the NeurIPS 2024 "Erasing the Invisible" competition, successfully removing watermarks from all 150 test images while maintaining exceptional quality preservation. The method demonstrated perfect watermark removal (A = 0) with minimal quality degradation (Q = 0.130). The findings of this study corroborate the efficacy of the proposed methodology, substantiating its capacity to meet numerous exacting quality assessment criteria. The investigation has demonstrated that the approach consistently achieves superior watermark removal effectiveness and represents a successful implementation of a removal attack against the StegaStamp algorithm.

The comprehensive evaluation across diverse metrics and the competition results collectively validate the robustness and effectiveness of the proposed watermark removal framework.